\documentclass{mn2e}
\usepackage{epsf}
\usepackage{graphics}

   \title[CB3 clumps]{The origin and structure of clumps along molecular outflows: the test case of CB3}

   \author[Viti et al.]{Serena Viti,$^{1,2}$\thanks{E-mail: sv@star.ucl.ac.uk}  Claudio Codella,$^3$ Milena Benedettini,$^1$ Rafael Bachiller$^4$ \\
$^1$CNR-Istituto di Fisica dello Spazio Interplanetario, Area di Ricerca di Tor Vergata, via del Fosso del Cavaliere 100, I-00133, Roma, Italy \\
$^2$ Department of Physics and Astronomy, University College London, Gower Street, London, WC1E 6BT, UK \\
$^3$ Istituto di Radioastronomia, CNR, Sezione di Firenze, Largo E. Fermi 5, 50125 Firenze, Italy \\
$^4$ IGN Observatorio Astron{\'o}mico Nacional, Apartado 1143, 28800 Alcal\`{a} de Henares, Spain }
\date{Received ; accepted }
\pagerange{\pageref{firstpage}--\pageref{lastpage}} \pubyear{2003}

\begin{document}

\label{firstpage}

\maketitle

\begin{abstract} 
\par
We investigate the origin of the small, chemically rich molecular
clumps observed along the main axis of chemically rich outflows such
as CB3 and L1157.  We develop a chemical model where we explore the
chemical evolution of these clumps, assuming they are partially
pre-existing to the outflow, or alternatively newly formed by the
impact of the outflow on the surrounding medium. The effects of the
impact of the outflow are reproduced by density and temperature
changes in the clump. We find that the observed abundances of
CH$_3$OH, SO and SO$_2$ are best reproduced by assuming a scenario
where the dense molecular gas observed is probably pre-existing in the
interstellar medium before the formation of their exciting
(proto)stars and that the clumpiness and the rich chemistry of the
clumps are a consequence of a pre-existing density enhancement and of
its interaction with the outflow.
\end{abstract}

%

\section{Introduction}

At an early stage in their evolution stars eject material in the form
of outflows. In fact, the first signs of an outflow are coupled with
infall motion and therefore with the first stages of star formation
(e.g. Bachiller 1996; Richer et al. 2000). Once the protostar is
formed, outflows are the main means of removing the material left over
from the collapse of the cloud.
\par 
Small molecular clumps ($\sim$ 0.1 pc) are detected in association
with several outflows; it is generally believed that the clumps are
generated by episodic mass loss of the forming object.  To date, there
is no detailed understanding of the role of the clumpiness in
outflows. In fact, two main kinds of clumps have been observed: (i)
high-velocity clumps or the so-called molecular bullets which are well
defined entities travelling at velocities larger than 100 km s$^{-1}$.
The prototype is L1448 (Bachiller, Mart\'{\i}n-Pintado \& Fuente
1991), where bullets appear in pairs with the members of a pair being
symmetric in both position and velocity with respect to the
star. These bullets are most likely associated with mini-bow shocks
formed by the outflow propagation (e.g. Dutrey, Guilloteau \&
Bachiller 1997). The molecular lines emitted by this kind of clump are
relatively weak, so their chemical composition remains unknown. (ii) A
second kind of clump, observed along a few outflows associated with
low- and intermediate-mass stars; these outflows stand out because of
their association with chemically rich clumps at definitely lower
velocity, such as L1157, BHR71, and CB3 (Bourke et al. 1997; Codella
\& Bachiller 1999; Bachiller et al. 2001).  The origin of these
chemically rich clumps is not yet clear.

In this paper, we investigate the origin and nature of the second kind
of clumps by the use of a chemical model that simulates the clump
formation and its subsequent interaction with the outflow. We consider
here two main scenarios: (1) pre-existing clumps, affected by the
outflow, and (2) newly-formed clumps, created by the outflow.  {\it
Note that we will use the word pre-existing to indicate a density
structure formed before the advent of the outflow.} Our definition
does not imply that the observed abundances (e.g Codella \& Bachiller
1999; Bachiller et al. 2001) are pre-existing to the outflow.
\par
In what follows, we introduce our two scenarios for the formation of
the clumps.

\begin{itemize}

\item[(1)] 
The clumps are pre-existing to the outflow; for example they may be
either remnant material of the collapsing parent cloud or completely
independent of the star formation process, but present homogeneously
in the dark molecular cloud (e.g. Falle \& Hartquist 2002; Morata et
al. 2003; Garrod et al. 2003). In this scenario, we observe these
clumps in association with molecular outflows because, as the outflows
travel through the molecular cloud, they interact with the clumps and
shock them. If this is the case, then the clumps may not be an
indication of the episodic nature of outflows.

\item[(2)] 
The cloud material is homogeneous and low in density but as an
episodic outflow forms, its interaction with the surrounding
homogeneous material will lead to a compression (increase in density)
and increase in temperature (with subsequent evaporation of the grains
mantles) in localized regions {\it only}, hence the observed
clumpiness (e.g Arce \& Goodman 2001, 2002, and reference therein). In
this scenario, the clumps are a direct manifestation of the episodic
nature of outflows.

\end{itemize}

Note that these scenarios do not necessarily exclude each other: in
fact, the morphology of the regions where outflows and jets propagate
have a very complicated geometry and structure by their very nature
(e.g Hester et al. 1998).
\par 
The main aim of this study is to investigate the nature of the
chemically rich clumps by modelling their chemical evolution and by
comparing the models with observations. In particular, we attempt to
identify observable species that can be used as discriminants between
the two scenarios. In this paper, we will focus our attention on the
clumps located in the intermediate-mass star forming region CB3
(Codella \& Bachiller 1999). In addition, a low mass case, represented
by the L1157 outflow, will be briefly discussed. Our model is
described in Section 2, and our results are presented, discussed and
compared with observations in Section 3. A brief conclusion is given
in Section 4.

\section{The Model}

\begin{figure*}
\vspace{8cm}
\includegraphics{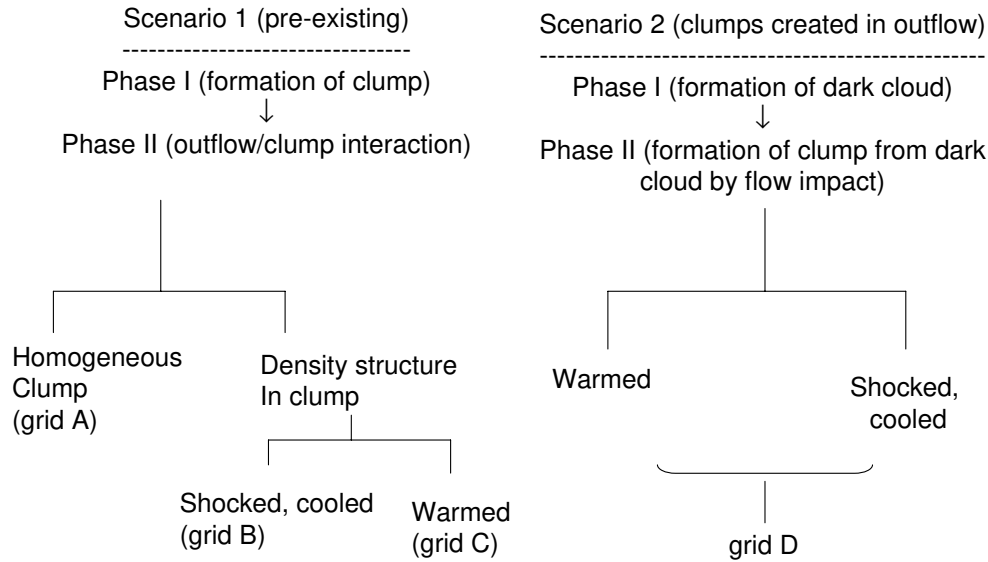}
\caption[]{A flow diagram of the two Scenarios as treated by the chemical model (see Section 2).}
\label{fg:scheme}
\end{figure*}

The aim of the modelling is to explore the chemistry in the different
scenarios of clump formation described above. The basic chemical model
we adopt is a modification of the time-dependent model employed in
Viti \& Williams (1999) and Viti et al. (2003).  The chemical network
is taken from the UMIST database (Millar et al. 1997; Le Teuff et
al. 2000). We follow the chemical evolution of 221 species involved in
3194 gas-phase and grain reactions. Our model is two-phase
calculation.  Phase I represents the formation of the pre-existing
clumps, Scenario 1, or of the homogeneous dark cloud, Scenario 2, from
diffuse gas, while Phase II represents the effect on the gas and dust
of the outflow. Figure~\ref{fg:scheme} shows a schematic of how the
two scenarios are treated in the model.

In Phase I, we allow the diffuse gas to collapse gravitationally
within the molecular cloud. During the collapse phase, gas-phase
chemistry and freeze-out on to dust grains with subsequent processing
are assumed to occur. The initial density of the gas is taken to be
100 cm$^{-3}$, and the final density is treated as a free parameter
(see later). During Phase I, the temperature is kept constant at 10
K. In Phase II we simulate the presence of the outflow (see Sections
2.1--2.3 for a detailed description of each grid of models) by an
increase in temperature and an assumed subsequent thermal or
(non-dissociative) shock-induced evaporation of the grain
mantles. Sputtering of the core of the grains and fast, dissociative
shocks are not included.  Phase II was halted at 10$^5$ yr.

In Phase II, Scenario 1, we have investigated both a uniform and
homogeneous clump (the set of physical parameters for which we have
explored the chemistry is here called Grid A), and clumps with a
density structure (Grids B and C). We have also considered the
possibilities that after the initial increase of the temperature (to
$\sim$ 100 K), the temperature remains constant (Grid C) or enters a
shocked phase to a higher temperature, followed by cooling (Grid
B). See 2.1--2.3 for more details.

\par In Scenario 2 the
outflow impacts on a dark molecular cloud, of density $\sim$ 10$^4$
cm$^{-3}$, and induces the formation of clumps. In this case we use
Phase I of the model to simulate the formation of the dark cloud, and
Phase II to simulate the formation of the clumps due to the impact of
the outflow. During Phase II, the gas and dust temperatures increase
as the clump forms, and the grains evaporate. Here too we investigated
the possibility that after the initial increase to 100 K, the
temperature enters a shocked phase to a higher temperature and
subsequent cooling. This scenario will be modelled in a range of
physical parameters here labelled Grid D.

Within each grid we have explored a reasonably large parameter space.
To impose some constraints to the initial parameters, we adopt the
physical parameters derived by Codella \& Bachiller (1999) for the
clumps observed along CB3, namely: a size for the final clump of
$\sim$ 0.12 pc, a final density of about 10$^5$--10$^6$ cm$^{-3}$, a
final temperature of 100 K, and an age of 10$^5$ yrs.

\subsection{Scenario 1: Grid A}

Table~\ref{tb:gridA} lists the details for the models computed for
Grid A.  Grid A consists of 9 models where, in Phase I, we vary the
following parameters: i) the type of collapse, ii) the depletion
efficiency, iii) the final density, and iv) the initial sulphur
elemental abundance. \par The collapse is either treated as free-fall
as described by Rawlings et al. (1992) or retarded. \par The depletion
efficiency is determined by what fraction of the gas phase material is
frozen on to the grains, and undergoes hydrogenation.  Several routes
of hydrogenation for the most significant species (O, N, C, CO) have
been explored.  The freeze-out fraction is arranged by adjusting the
grain surface area per unit volume, and assumes a sticking probability
of unity for all species. The fraction of material on grains is then
dependent on the product of the sticking probability and the amount of
cross section provided per unit volume by the adopted grain size
distribution. This product was varied so that at the end of Phase I,
we would have different percentages of ices (see
Table~\ref{tb:gridA}). Note that as the chemistry is time-dependent,
different species form at different times and as a consequence the
material frozen out on the grains at any one time is $not$
representative of the whole gas but of selected species (in this case,
we chose to monitor frozen CO). \par For the final density, we chose a
lower limit of 10$^5$ cm$^{-3}$ and an upper limit of 5$\times$10$^6$
cm$^{-3}$. \par The initial relative abundance of sulphur is very
uncertain (e.g., Ruffle et al. 1999); we chose the solar value as an
upper limit and a factor of hundred lower than the solar value as a
lower limit. During Phase II, the temperature rises fast and it
reaches a maximum of 100 K.

\begin{table}
\caption{Model Parameters for Grid A. The notation a(b) signifies
a $\times$ 10$^{b}$. The model number is listed in Column 1; Column 2
shows the density of the gas; Column 3, 4 and 5 shows respectively the
percentage of the gas depleted onto grain at the end of Phase I, the
type of collapse (B=1 free--fall; B=0.1 retarded), the initial sulphur
abundance.}
\begin{tabular}{ccccc}
\hline
Model & $n_f$ (cm$^{-3}$) & Depletion (\%) & B & Sulphur \\
\hline
A1 & 1(5) & 15 & 1.0& 1.3(-5) \\
A2 & 1(5) & 35 & 0.1& 1.3(-5) \\
A3 & 1(5) & 11 & 0.1& 1.3(-5) \\
A4 & 1(5) & 20 & 1.0& 1.9(-6) \\
A5 & 1(6) & 20 & 1.0& 1.3(-5) \\
A6 & 1(6) & 30 & 1.0& 1.3(-7) \\
A7 & 1(6) & 55 & 1.0& 1.3(-7) \\
A8 & 1(6) & 80 & 1.0& 1.3(-7) \\
A9 & 5(6) & 25 & 1.0& 1.3(-5) \\
\hline 
\end{tabular}
\label{tb:gridA}
\end{table}

\subsection{Scenario 1: Grids B and C}

\begin{figure*}
\vspace{8cm}
\includegraphics{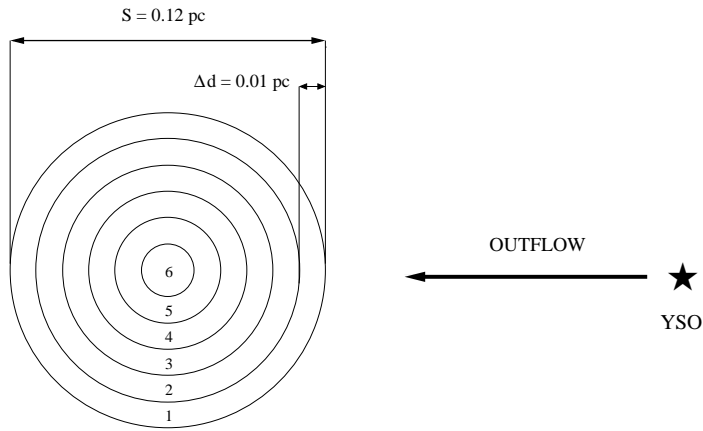}
\caption[]{A schematic picture (not to scale) of the inhomogeneous clump modelled in Grids B, C and D. The chemical model is in 1-D, hence the chemistry is determined only along one symmetry axis. The geometry is then taken into consideration when estimating the column densities.}
\label{fg:drawing}
\end{figure*}

In Grids B and C we assumed that the 0.12 pc final clump is
inhomogeneous
(See Figure~\ref{fg:drawing}).
In Phase I, every depth point starts at 100 cm$^{-3}$ and collapses
until it reaches a final density, which varies from edge to centre of
clump, according to the density law derived 
by Tafalla et al. (2002) for starless cores (and which we adopt): 
\\

\begin{equation}
{\rm n(r)} = {\rm n}_{{\rm o}}/{\rm [1+(r/r}_{\rm o}]^{\alpha})
\end{equation}

where r$_{\rm o}$ and ${\rm n}_{{\rm o}}$ are the largest distance
from the centre of the core and the peak density,
respectively. $\alpha$ is the asymptotic power index and it is taken
to be $\sim$ 2.5 (see Table 2 in Tafalla et al. 2002). We divided our
clump in 6 shells differing in density ($n_f$), visual extinction
(A$_V$), and distance from the (future) outflow.
Table~\ref{tb:shells} shows the parameters for each of the shells and
Figure~\ref{fg:drawing} gives a simplified graphical representation of
the clump/outflow system (not to scale). In Table~\ref{tb:shells},
Column 2 lists the distance of the shell from the edge, and Column 3
lists the final density for each shell.  Phase I is halted when the
final density is reached for each point.

\par
Table~\ref{tb:gridsB_C} shows the details of the several models
belonging to Grids B and C.  To start with, we ran Phase I for two
different percentages of gas as ices in the mantles (B1 and B2 models
and C1 and C2 models). B3 and B4 are similar to respectively B1 and B2
but we adopted a variation to the routes of hydrogenation for some
species (See Section 3.3.2 for details).

In Phase 2, Grid B simulates the arrival of the outflow by allowing
the clump to undergo a non-dissociative shock where the mantle of the
grains is evaporated and where the temperature of the gas reaches 1000
K and stays at this temperature for a short period of time (100 yr or
so, following Bergin et al. 1999), after which the clump cools down to
100 K. In Grid C, the arrival of the outflow is simulated by simply an
increase of gas and dust temperature up to 100 K, with a subsequent
evaporation of the grains which occurs either instantaneously (C1--C2)
or via time-dependent evaporation (C3), as in Viti \& Williams (1999).
Both Grids were run for 10$^5$ years.

\begin{table}
\caption{Density structure for all Grids, apart from Grid A. The notation a(b) signifies
a $\times$ 10$^{b}$. $d$(pc) is the distance between the mid-point of each shell and the edge of the clump (see Figure 1). n$_f$ is the final density. A$_V$ is the visual extinction between the outflow and the mid-point of each shell.}
\begin{tabular}{cccc}
\hline
Shell & $d$ (pc) &  n$_f$ (cm$^{-3}$) & A$_V$ (mags) \\
\hline
1 & 0.005 & 5.0(5) & 10 \\
2 & 0.015 & 6.4(5) & 25 \\
3 & 0.025 & 7.8(5) & 45 \\
4 & 0.035 & 9.0(5) & 70 \\
5 & 0.045 & 9.8(5) & 95 \\
6 & 0.055 & 1.0(6) & 116 \\
\hline
\end{tabular}
\label{tb:shells}
\end{table}

\par

\begin{table}
\caption{Model Parameters for Grids B and C. The notation a(b) signifies
a $\times$ 10$^{b}$. The model number is listed in Column 1; Columns 2,
3, 4 and 5 indicate whether the clump is pre-existing, the percentage
of freeze-out, whether the gas is shocked, whether the evaporation has
occurred instantaneously (I) or not (TD), respectively. The last
column indicates whether a fraction of frozen H$_2$CO is converted into
methanol - see Section 3.3.2 for details.}

\begin{tabular}{ccccc}
\hline
Model & FR (\%) & Shock & Evaporation & H$_2$CO $\Rightarrow$ CH$_3$OH \\
\hline
B1 & 40 & yes& I & no\\
B2 & 60 & yes& I & no \\
B3 & 60 & yes& I & yes \\
B4 & 80 & yes& I & yes\\
C1 & 40 & no & I & no \\
C2 & 60 & no & I & no \\
C3 & 60 & no & TD & yes \\

\hline
\end{tabular}
\label{tb:gridsB_C}
\end{table}

\subsection{Scenario 2: Grid D}
In Scenario 2, the formation of molecular clumps along a molecular
outflow occurs as a direct consequence of the impact of the outflow on
a dark molecular cloud. We adopt the density structure of Eq. 1
(Tafalla et al. 2002) for this Scenario. Phase I simulates the
formation of a a single-point dark core from a diffuse medium (n = 100
cm$^{-3}$) collapsing in free-fall until n= 10$^4$ cm$^{-3}$ is
reached. The chemistry arising in Scenario 2 is explored for a range
of physical parameters listed as Grid D (Table~\ref{tb:gridD}. We
chose the freeze-out parameter so that the abundances of key species
at the end of the collapse were consistent with observations of dark
clouds (van Dishoeck 1998; cf. Bergin et al. 2002 for the CO
depletion). In Phase II we simulate the arrival of the outflow and the
formation of high density clumps with final densities as in
Table~\ref{tb:shells}. We take as free parameters the outflow velocity
and the initial distance of the dark cloud to the outflow. The models
we have run are listed in Table~\ref{tb:gridD}; the ranges of
velocities and distances were determined by considerations of the
values derived for CB3 (Codella \& Bachiller 1999). Note, however,
that a high velocity outflow at a too close initial distance to the
gas material (Models D2, D3, D6, D9) do not give the clump the time to
form and reach the high densities considered.  The density and
temperature vary in the following way:

\begin{equation}
{\rm n(r) = n}_{i}{\rm (r/r}_i)^{-3/2}
\end{equation}

and 

\begin{equation}
{\rm T(r)= T}_o{\rm (r/r}_o)^{-0.4}
\end{equation}

where r$_i$ is the initial distance, T$_o$ is an estimate of the
hottest gas ($\sim$ 100-200 K) and r$_o$ is the closest (final)
distance between the outflow and each shell within the clump that we
are modelling, such that at r = r$_o$, T(r) = T$_o$. The initial
distance varies of course from shell to shell; the distance between
the edge of the clump (shell 1) and the outflow for each model is
listed in Table~\ref{tb:gridD}.  The temperature varies with time (and
therefore with distance) as in Equation 3 (see Rowan-Robinson 1980).
\par Model D1 was also run with the addition of a post-shocked phase (with a
temperature of 1000 K for 100 yrs, followed by cooling) during the formation of
the clump, as in Grid B.
\begin{table}
\caption{Model Parameters for Grid D. The model number is listed in Column 1; Column 2 and 3 are the velocity of the outflow and the initial distance between the cloud and the outflow, respectively. }
\begin{tabular}{|c|cc|}
\hline
Model & Velocity (km/s) & $r$ ( pc) \\
\hline
$^*$D1 & 2 & 0.1 \\ 
D2 & 20 & 0.1 \\
D3 & 100 & 0.1 \\
D4 & 2  & 0.2 \\ 
D5 & 20 & 0.2 \\
D6 & 100 & 0.2 \\
D7 & 2 & 0.4 \\
D8 & 20  & 0.4  \\ 
D9 & 100 & 0.4 \\
\hline
\multicolumn{3}{|l|}{$^*$Ran with two different temperature behaviours. See Section 2.3.} \\ 
\end{tabular}
\label{tb:gridD}
\end{table}

\section{Results}

In Section 3.1 we give results from Grid A and briefly compare them to
the observations. In Section 3.2 we present our results from Grids B,
C and D while in Section 3.3 we compare them qualitatively with
observations of the clumps detected along CB3 and L1157.

\subsection{Grid A}

\begin{figure*}
\vspace{12cm}
\includegraphics{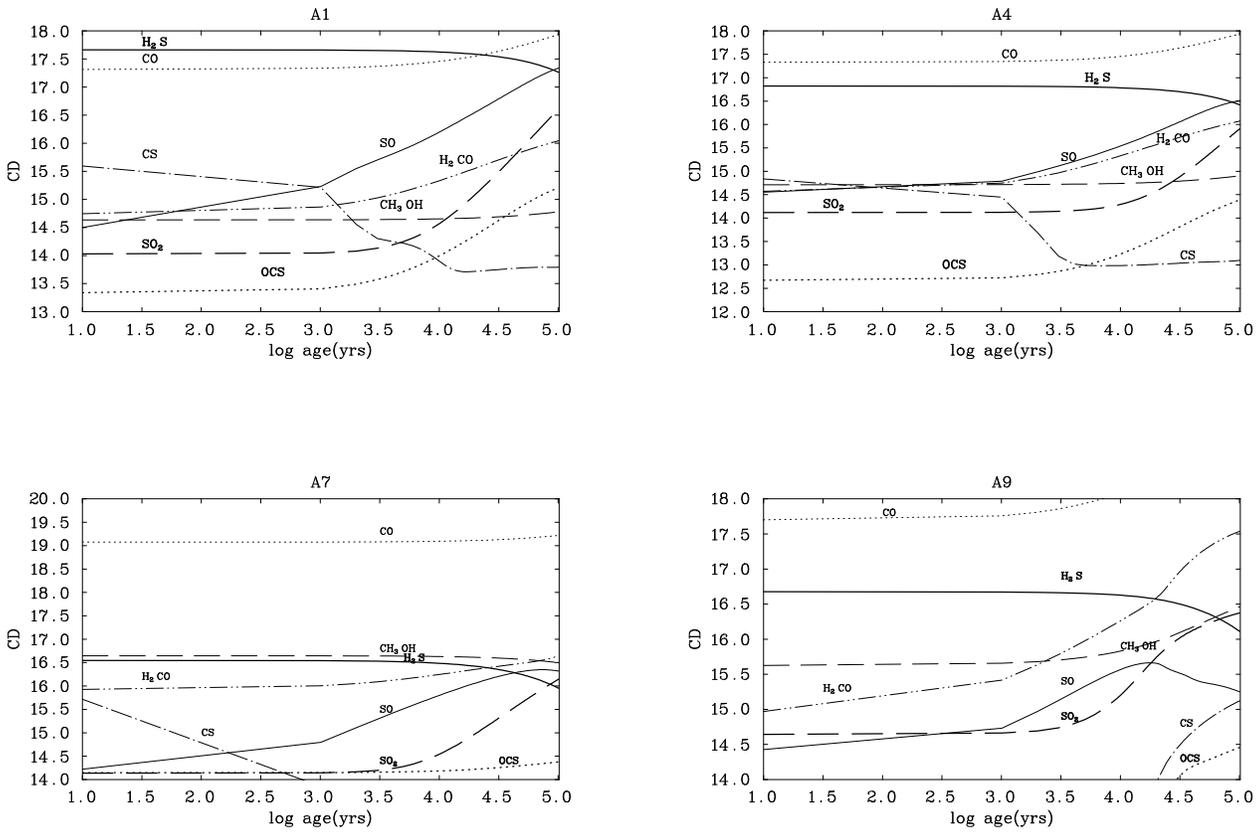}
\caption[]{The column densities (in cm$^{-3}$) of a selection of species relative to
hydrogen over time for selected models from Grid A.}
\label{fg:grida}
\end{figure*}

In Table~\ref{tb:cdA} we list the column densities at 10$^5$ yr,
estimated from some of the models of Grid A as compared to the
observations while Figure~\ref{fg:grida} shows the column densities of
selected species as a function of time for some models.

\begin{table*}
\caption{Column densities (in cm$^{-2}$) of selected species for some models from Grid A versus the column densities observed in CB3 (Column 6).  }
\begin{tabular}{|c|cccccc|c|}
\hline
& A1 &   A4  & A6   & A8   & Obs\\
\hline
CS & 6.2(13) &   1.2(13) &   9.8(13) &   1.1(14) &   1.7(14) \\
H$_2$CO & 1.1(16)  & 1.2(16)   & 3.5(16)   & 1.4(17)   & 5.9(14) \\
H$_2$S & 1.8(17)  & 2.6(16)   & 1.1(18)   & 1.4(16)   & 1.9(14) \\
SO$_2$ & 4.1(16)  & 8.1(15)   & 1.1(19)   & 2.8(16)   & 5.6(14)  \\
OCS & 1.6(15) & 2.5(14)   & 8.1(17) &   1.6(14)   & 9.8(13)  \\
SO & 2.2(17) &  3.2(16)   & 1.0(19) &   3.8(15)   & 1.3(14)  \\
CO & 8.7(17) &  8.7(17)   & 1.4(20) &  9.8(18)   & 2.0(17) \\
CH$_3$OH & 6.0(14) &  8.0(14) &   2.0(17)   & 2.9(16)   & 1.1(16) \\
\hline
\end{tabular}
\label{tb:cdA}
\end{table*}

From the table it is clear that the models from Grid A do not fit the
observations.  This is probably a consequence of the
oversimplification introduced by the one-density component.

Nevertheless, general trends do emerge from this simple grid of models:
\begin{enumerate}
\item a depleted sulphur initial abundance, rather than a solar one, is
preferred; we note from our results that, apart from CS (whose
abundance is mainly affected by the degree of depletion) other sulphur
species, such as SO, are far too abundant if the initial sulphur is
solar. This confirms previous studies (Ruffle et al. 1999; Viti et
al. 2003). \item Freeze-out on to grains must be effective at the
densities considered here; we believe that at least half of the gas is
depleted on to grain by the end of Phase I. \item None of the species
seem to be particularly affected by the type of collapse employed; for
simplicity we will then adopt a free-fall collapse for the remainder
of our Grids.
\end{enumerate}
So far, we looked at the column densities at 10$^5$ yr, a kinematical
age estimated by Codella \& Bachiller (1999) for the clumps along
CB3. However, this estimate may be easily off by over one order of
magnitude, since it strongly depends on the assumed geometry of the
outflow and it can thus vary in the 2 10$^4$ - 5 10$^5$ yr range
(Codella \& Bachiller 1999).

From Figure~\ref{fg:grida}, we can see that certain species, such as
CS, H$_2$CO, and SO$_2$ are very time-dependent; so for example the
SO$_2$ column density is close to the observed one if the clump is
less than 10$^4$ yr old, while CS is best matched at later times (see
Model A9); if the density is closer to 10$^6$ cm$^{-3}$ and sulphur
initial abundance is low (models A7 and A9 in Figure~\ref{fg:grida})
SO is reasonably matched before 3000 yr. These brief considerations
underline the importance of constructing a clump with a density
structure since not only we notice that different species match
observations at different densities, but also the degree of freeze out
(which determines a great deal of the time dependent chemistry) is
density dependent.

\subsection{Grids B, C and D}

As these grids share the same density structure, we discuss their
trends together.  Details of the parameters employed for grids B, C
and D are listed in Tables 3 and 4 while the column densities for most
models at 10$^5$ yr are listed in Table~\ref{tb:gridsb_c_d}. Before
commenting on whether or not they compare well with the observations
(see Section 3.3), we look at the general trends as in the previous
section.

{\small
\begin{table*}
\caption{Column densities (in cm$^{-2}$)  for selected models from Grids B, C, and D at $\sim$ 10$^5$ yr vs Observed column densities}
\begin{tabular}{|c|ccccccc|c|}
\hline
&  B1 & B2 & B3   & C1 & C2   & D1  &D8 & Obs  \\
\hline
CS & 1.66(15) & 1.24(15)& 4.76(15) &   5.79(13) &1.60(14)&  8.98(15) &   9.64(15)& 1.7(14)  \\
H$_2$CO  &2.32(17)&2.64(17)& 1.73(17) &   4.24(16)&1.20(17)& 2.78(15)&  2.74(15)&5.9(14) \\
H$_2$S & 5.94(14) &5.97(14)& 2.52(14) &   4.00(15) &1.50(15)  &4.90(13)   &1.15(14)&1.9(14) \\
SO$_2$ & 5.64(15) &5.93(15)& 2.59(15) &   3.82(15) &6.56(15) &1.15(13)  &2.13(13)& 5.6(14) \\
OCS &  1.61(14) & 1.95(14)& 6.15(13) & 2.10(13) &2.71(13)& 2.76(14)  &2.77(14)&9.8(14) \\
SO &  2.74(13) & 3.21(13)& 1.34(13) &  8.89(14) &6.74(13)& 2.43(13)&  7.79(12)& 1.3(14)\\
CO &  1.11(18) &9.40(17)& 1.16(18) &  2.17(18) &1.56(18)&6.15(18) &   6.32(18)&2.0(17) \\
CH$_3$OH  & 6.05(15) &6.40(15)& 1.22(16) &   5.48(15)&6.20(15)&2.12(14)&2.62(14)& 1.1(16) \\
\hline
\end{tabular}
\label{tb:gridsb_c_d}
\end{table*}
}

\subsubsection{Grid B vs. C} 

Table 6 shows that varying the percentage of gas as ices in mantles in
Grid B (e.g B1 vs. B2) does not significantly affect the abundances of
most species, while in Grid C the abundance of sulphur-bearing species
can vary by up to one order of magnitude.
\par More significant differences are found between models of
Grid B and those of Grid C, especially among the sulphur bearing
species; this is not surprising as sulphur-bearing species are known
to be good tracers of high temperature gas (e.g, Hatchell \& Viti
2002; Hatchell et al. 1998). We find that CS is as much as two to
three orders of magnitude more abundant in Grid B than in Grid C
models while H$_2$S, SO and OCS vary by two orders of magnitude at
most; H$_2$S is the driving species for the sulphur chemistry: it is
enhanced on the grains during freeze out (via hydrogenation of
sulphur) and it is then evaporated and dissociated during Phase II;
since it is easily dissociated at high temperatures, it is, of course,
more abundant in Grid C models (where the absence of a high
temperature phase slows down its dissociation). A faster dissociation
of H$_2$S in Grid B increases the abundance of other sulphur species,
in particular we note OCS (an otherwise underabundant
species). SO$_2$, on the other hand, does not seem to be affected as
much. \par We conclude that, if the clump is pre-existing, then we
should be able to easily discern weather it has undergone a high
temperature phase or not using sulphur-species as tracers: moreover,
if we assume that once the outflow has reached the clumps,
non-dissociative shocks must occur, then sulphur bearing species can
indeed be used as chemical clock in order to determine the age of the
outflow/clump system, as proposed also by Codella
\& Bachiller (1999) and Bachiller et al. (2001) for the high-velocity 
outflowing gas.

\subsubsection{Grids B, C vs. D} 
As expected, the differences between Grids B, C, and Grid D involve
more species. \par In particular, we find that, apart from sulphur
bearing species, CH$_3$OH is a good tracer of the different
scenarios. CH$_3$OH is not easily made in the gas-phase and is thought
to be mainly formed on the grains via hydrogenation of a fraction of
CO (Millar \& Hatchell 1998) and possibly H$_2$CO. In a high
temperature environment, CH$_3$OH has an alternative route of
formation via water; in fact, we note from Table~\ref{tb:gridsb_c_d}
that, at equal freeze out, methanol is higher in B than in C models
(cf B1 and C1), although at high freeze out they are comparable (cf B2
ad C2). This implies that a high abundance of CH$_3$OH depends more
strongly on the amount of CO on the grains than on whether a high
temperature phase has occurred. \par For the same reason, Grid D
models have a low abundance of CH$_3$OH as the formation of the clump
took relatively little time, hence depletion and hydrogenation onto
the grains was not very effective (Rawlings et al. 1992). If a high
temperature phase is included for D1, the methanol abundance does
indeed increases, but all the sulphur-bearing species increase by over
one order of magnitude.
\par
Methanol and sulphur-bearing species may therefore be the ideal
candidates to determine whether the clumps originate as a consequence
of the outflow or whether they are, at least partially, pre-existing.

\subsection{Comparisons with observations}

\subsubsection{Derived gas parameters for CB3}

In this section we briefly summarise how the gas parameters of the CB3
outflow have been derived from the IRAM 30-m observations (Codella \&
Bachiller 1999). For further details we refer the reader to the
Codella \& Bachiller (1999) paper.
\par 
The molecular outflow in CB3 is associated with at least four clumps
located along the main axis. For the sake of clarity we calculated and
used averaged values.  The column density of SO, as well as the
kinetic temperatures ($T_{\rm kin}$) and the hydrogen density ($n_{\rm
H_2}$), have been derived by means of statistical-equilibrium LVG
calculations. These two species, together with CH$_3$OH, have been
observed in three emission lines. For CH$_3$OH a rotation diagram has
been used to estimate the rotational temperature and its column
densities. Gas density estimates have also been derived from the
methanol emission patterns by measuring line intensity ratios. The
column densities for the other molecules, observed in a single
transition, have been estimated by assuming the lines optically thin
and in LTE conditions, and by using standard partition functions. In
this case, a temperature of 100 K has been adopted following the LVG
indications for the clumps: the temperature uncertainty yields to an
accuracy of the derived values within a factor of 10.

\par Note that
with LVG calculations it is not possible to separate the effects on
the excitation of the density and temperature. Moreover, LVG
calculations do not take into account any correction due to the
different beam filling factors at the three wavelengths of the
observed transitions. As discussed by Codella \& Bachiller (1999), if
the source sizes were to be definitely smaller than the three beam
widths, i.e. a sort of point-like source, the LVG code would lead to
an overestimate of the excitation conditions: $T_{\rm kin}$ and
$n_{\rm H_2}$ would be reduced by factors of about 2, while the column
densities by a factor of 9.
\par 
The kinematical age of the outflow is estimated to be between
2$\times$10$^4$--5$\times$10$^5$ yrs, and it has been derived by
comparing the positions of the farthest clumps, with respect to the
driving source, assuming that the material travelled from the centre
to its present location with a typical observed velocity of $\sim$ 2
km s$^{-1}$, and correcting for the projection effect, given an
inclination to the plane of sky of 30$^{\rm o}$.

\subsubsection{Model comparisons with CB3}

Table~\ref{tb:gridsb_c_d} compares the column densities of Grids B, C
and D with the observed values derived from the observations of
Codella \& Bachiller (1999), at 10$^5$ yr. Note that, unlike in the
case of Grid A, the column density is calculated adding the
contribution of each shell of the clump using the following formula:

\begin{equation}
{\rm N}_i =  \sum_{s} (X_{i} \times {\rm L} \times {\rm n}_s \times f_{i,s})
\end{equation}

where $s$ is an index indicating the shell, $X_i$ is the fractional
abundance of species $i$, L is the length of each shell, n is the
density of the shell, $f_{i,s}$ is the weighted beam dilution. In
Grids B, C and D we do not need to approximate the inhomogeneity of
the medium (as in Grid A) by multiplying the density by the visual
extinction.
\par
Also, since the age of the observed gas is uncertain (see Section
3.3.1), we also show in Figure 4 the behaviour of selected species as
a function of time for some of the models listed in
Table~\ref{tb:gridsb_c_d}. The chemistries of some species, such as,
in fact, the ones which show most discrepancies with the observations,
are highly time dependent; for example, H$_2$CO, and some sulphur
species, can be almost two orders of magnitude lower in abundances at
early times.

\par
At first sight, from Table~\ref{tb:gridsb_c_d}, it appears that none
of the models succeeds in reproducing $all$ the observed column
densities within the observed uncertainty (taken to be one order of
magnitude, see Section 3.3.1).

\begin{figure*}
\vspace{12cm}
\includegraphics{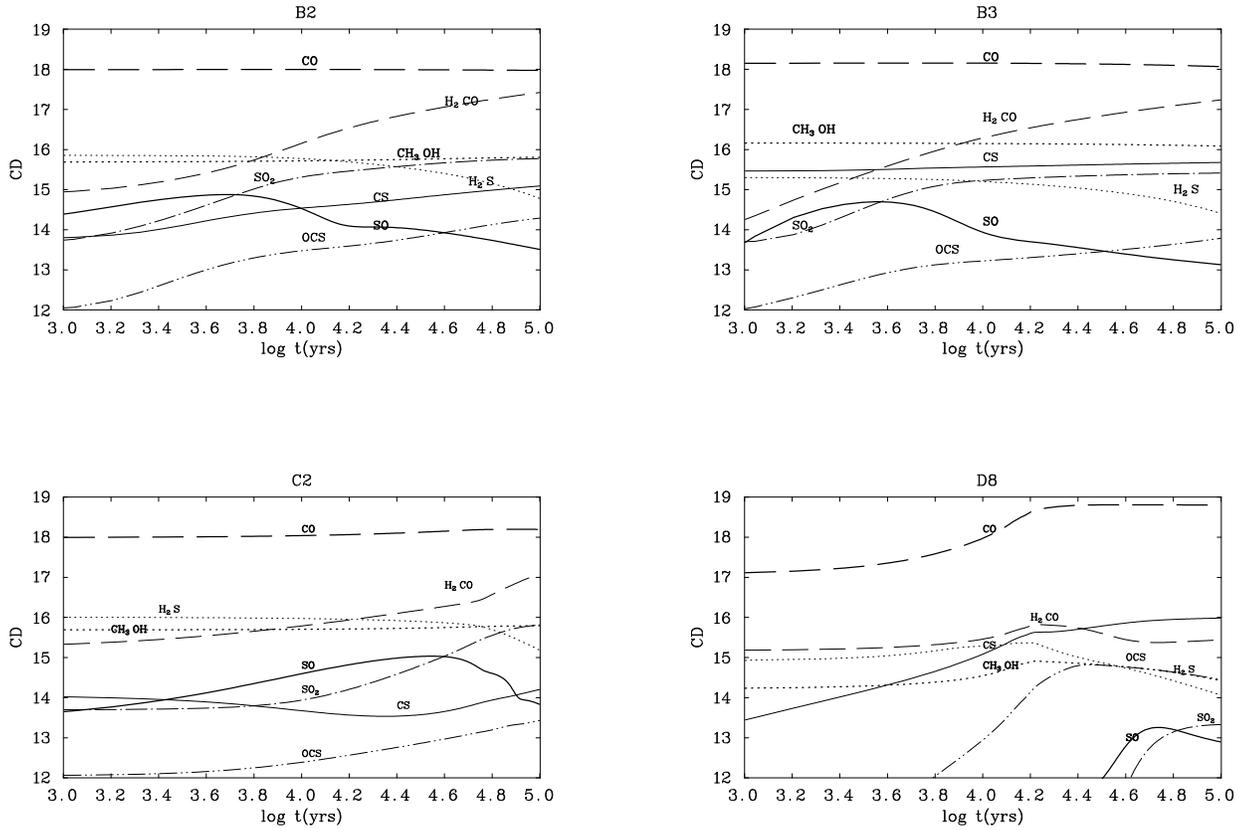}
\caption[]{The column density of a selection of species over time for selected models from Grid B, C and D.}
\label{fg:time1}
\end{figure*}

\begin{figure*}
\vspace{12cm}
\includegraphics{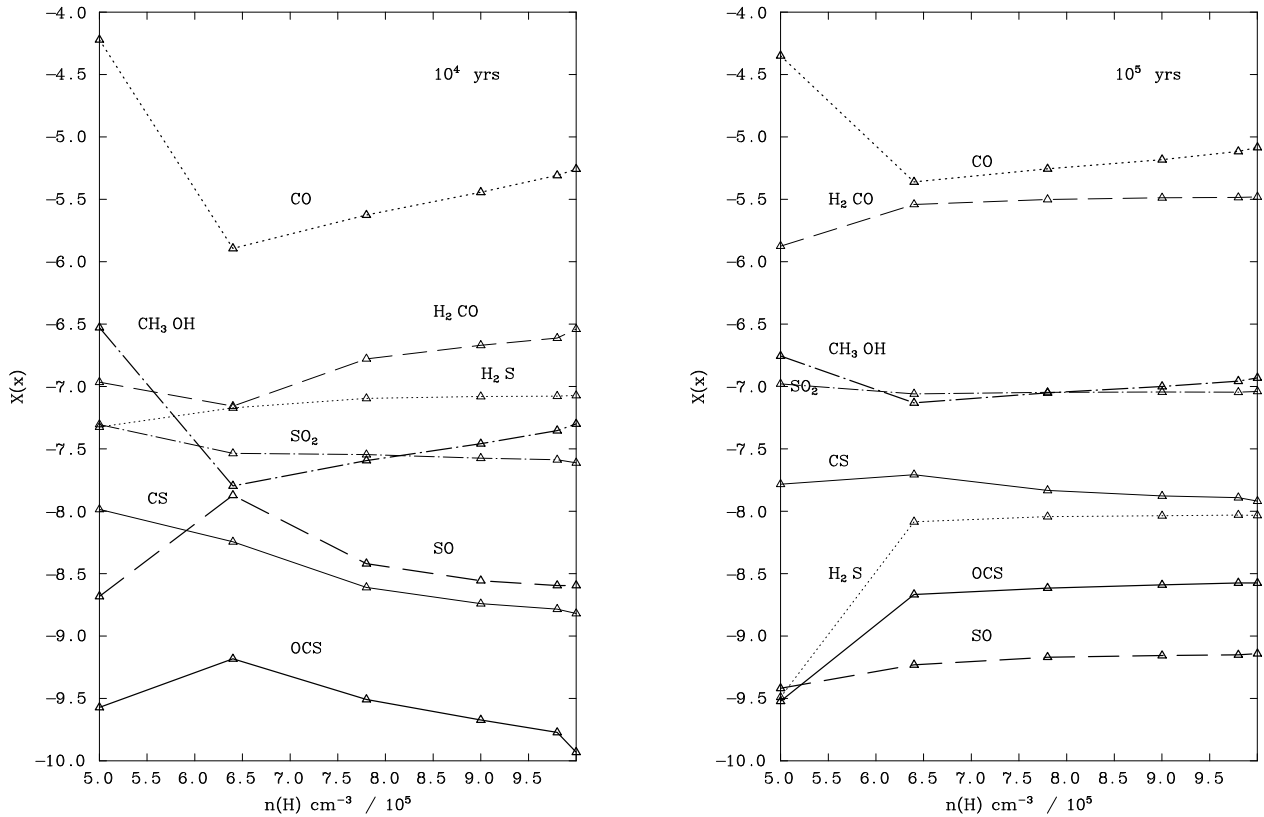}
\caption[]{Fractional abundance of selected species as a function of density for
 Model B2 at 10,000 yr (left) and 10$^5$ yr (right). The diamond marks indicate the densities at which each shell was computed. The points are joined for clarity. }
\label{fg:emission}
\end{figure*}

However, we can see that Grid D fails to reproduce important species
such as methanol and sulphur species and it is therefore the worse
matching grid; Grid C improves over Grid D but still fails, in
general, to match the observations for some sulphur species and, to a
lesser extent, for methanol (see Section 3.2.2).
\par
A first conclusion that can be made is therefore that {\it the
chemically rich clumps along CB3 can not be formed completely by
compression of the gas due to the advent of the outflow.} This, of
course, implies that the high density of the clumps must have been
reached before the advent of the outflow. This preliminary conclusion
seems to be in contradiction with the conclusion by Arce \& Goodman
(2001, 2002) that the outflow clumps are the product of the sweeping
up ambient gas. However, their assumption is that the ambient gas is
homogeneous, at a density of $\sim$ 5$\times$10$^3$ cm$^{-3}$, and
that the outflow will sweep-up a volume of $\sim$ 0.5 M$_{\odot}$,
$maintaining$ the same density of pre-existing cloud material - so,
effectively, from a chemical point of view, the clump is indeed
pre-existing the outflow/star system as the final density was reached
without the intervention of the outflow. In this picture the
morphology of the swept-up ambient gas and $not$ its high density is a
manifestation of the episodic nature of the outflow.
\par

Grid B models seem to be the best, in particular B2 gives the closest
match with observations, within half an order of magnitude (well
within the observed error bars, see Section 3.3.1): apart from CO, all
species are well matched for t$\ge$ 10,000 yr, although H$_2$CO is
definitely best matched at t $\le$ 10,000 yr. Although not obvious
from Figure 4 (but evident from Table~\ref{tb:gridsb_c_d}), the CO
abundance in Grid B is $\sim$ half the one in Grid C, due to the
presence of a high temperature gas phase. However this difference did
not increase by increasing the high temperature phase to 500 yr
(models not shown), as opposed to 100 yr as assumed in Grid B.
\par
In general, we find that the theoretical abundance of H$_2$CO is
overabundant with respect to the observations. One of the formation
routes for H$_2$CO is the hydrogenation of CO, HCO and HCO$^+$ on the
grain mantles.  We tried a model, B3 (also shown in
Figure~\ref{fg:time1}), similar to B2 but where i) frozen H$_2$CO is
converted into methanol (a possible hydrogenation), ii) frozen HCO and
HCO$^+$ remain unaltered, upon depletion, iii) none of the carbon that
freezes onto the grains hydrogenate into methane (unlike in previous
models, where a percentage of carbon atoms became CH$_4$); the latter
reaction is important because some of the H$_2$CO formed during the
warm phase is formed by the reaction of water with CH$_3$ that comes
from the dissociation of CH$_4$. We also computed a model, similar to
B3, but with higher freeze out (B4, not shown in
Figure~\ref{fg:time1}). Although the final (at 10$^5$ yr) fractional
abundance of H$_2$CO is lower than in B1 or B2, we notice that in B3
and B4 the CS abundance is too high.

\par
We can conclude therefore that the most likely scenario of the clumps
observed along the outflow is a scenario where {\it the clumps are
pre-existing, but undergo a high temperature phase (caused by a
non-dissociative shock) when the outflow arrives, which leaves them
altered in temperature for a period not longer than 5$\times$10$^4$
yr}.

\subsubsection{The physical structure of the CB3 clumps}
There are several reasons why Grid B models do not always
quantitatively match the observations: for example, so far we have
only looked at the total column density as coming from the whole
clump. However, as shown from the results of Grid A, we know that the
clumps must be formed by several density components; the density
structure adopted here is believed to be representative of low-mass
star-less cores (Tafalla et al. 2002) but it is likely that the
density profile for the star-less, small clumps considered here have a
different structure. It is therefore worth making some general
considerations on the possibility that different species are emitted
from different components of the gas.
\par Figure~\ref{fg:emission} shows, at two different epochs, the
fractional abundances of selected species as a function of the density
for the model B2. From this figure it is clear that if the clumps are
as old as 10$^5$ years the emission comes effectively only from a
two-components density structure, one at $\sim$ 5$\times$10$^5$
cm$^{-3}$ and the other at $\sim$ 10$^6$ cm$^{-3}$. It may be that
some gas comes from a lower density component but our `edge' (lower)
density was determined by imposing a upper density of 10$^6$ cm$^{-3}$
(as derived by observations, see above). In fact, if we look at CO it
seems as it is indeed the lower density component that is inconsistent
with the observations. Note that although from this figure it seems as
an homogeneous gas at 10$^6$ cm$^{-3}$ would have matched the
observations, this is not the case at earlier times, and overall
during the collapse phase (which determines the grain surface
chemistry), when the chemistry is much more density dependent.
\par A
possibility is that the clumps observed are smaller than the size
implied by the observations (and in fact 0.12 pc is an upper limit),
but at higher density: this would imply a smaller A$_V$ and therefore
smaller column densities for the highest density components. General
trends that can be derived from Figure~\ref{fg:emission} (overall left
panel) are that:
\begin{enumerate} 
\item CO and CH$_3$OH,
and to a lesser extent CS and SO$_2$, are mainly emitted by the lower
density components.
\item H$_2$CO is slightly more abundant in the highest
densities parts.
\item H$_2$S is more or less constant at early times, while SO and OCS
seem to be highest at intermediate densities. At 10$^5$ years, H$_2$S
is mainly emitted from the highest density component.
\end{enumerate}
Finally, it is not excluded that each clump observed by Codella \&
Bachiller (1999) does in fact contain substructures, not resolved in
the single-dish observations because of the large distance of the source.
\par

\subsubsection{Another test case: L1157}

Here we briefly compare our models with the chemical properties of the
clumps observed in L1157.  The associated outflow is driven by a
low-mass YSO and will be subject of a future detailed study.
\par
Bachiller et al. (2001) have mapped the bipolar outflow in several
molecular emission lines and found chemical differentiation along the
outflow. In particular they detect small ($\sim$ 0.04 pc) clumps in
the southern lobe (B0--B2) which differ in densities and
temperatures. If the origin of these clumps is similar to those
observed along CB3, then their smaller size supports the possibility
that the clumps along CB3 are smaller than the size implied by the
observations due to the distance of the outflow (L1157 is only 440 pc
away while CB3 is at 2500 pc). The best defined clump, called B1, has
a density of 3--6$\times$10$^5$ cm$^{-3}$ and a kinetic temperature of
$\sim$ 80 K, so, apart from its size, it is comparable to the clumps
along CB3 and can therefore be briefly compared with our model
results. Table~\ref{tb:l1157} compares the column densities,
recalculated (as if coming from a smaller size) with Equation 4, of
our best-matching model (B2) at 10$^5$ yr and of D2, with the derived
column densities. Table~\ref{tb:l1157} clearly shows that the match
between B2 and the observations improves as the size of the clump is
reduced. Grid D is still unable to match the methanol and most of the
sulphur-bearing species confirming the results obtained for CB3.

\begin{table}
\caption{Column densities (in cm$^{-2}$) of selected species for B2 and D2 versus the observed column densities of the L1157 clumps.  }
\begin{tabular}{|c|ccc|}
\hline
 & B2 & D2 & L1157 \\
\hline
CS &1.15(14)& 3.07(15)  & 2.7(14) \\
H$_2$CO &  4.96(15) &8.57(14)& 3--8(14)  \\
H$_2$S &  1.94(15) &5.53(13) & 3.9(14)  \\
SO$_2$ &  6.92(14) & 6.40(12) &3.0(14) \\
OCS &  1.00(13) & 1.15(14)& 5(13) \\
SO &  1.04(14) & 3.64(12) &3.5(14) \\
CO &  3.28(17) & 2.10(18) & 1.4(17) \\
CH$_3$OH &  1.72(15) & 1.17(14) & 0.5-2.6(15) \\
\hline
\end{tabular}
\label{tb:l1157}
\end{table}

\section{Conclusions}

We have presented here a detailed time-dependent chemical model of the
chemically rich clumps observed along outflows.  This preliminary
study was aimed at finding some observable tracers that could help us
understanding the origin of the clumps with respect to the outflow. We
also attempted a qualitative comparison of our models with some
observations, in particular with the clumps observed along the CB3
outflow (Codella \& Bachiller 1999). Despite large uncertainties in
the mode of formation and several chemical assumptions we made, we
believe that we are able to constrain some of physical and chemical
parameters of the CB3 clumps.  Our conclusions are:

\begin{enumerate}
\item[1.] 
The initial sulphur abundance of the gas forming the clump can not be
solar. We find a depletion factor of $\sim$ 100, confirming the
findings of other studies (Oppenheimer \& Dalgarno, 1974; Ruffle et
al. 1999).
\item[2.] 
A substantial freeze out must occur during the formation of the clump,
regardless of its mode of formation.
\item[3.] 
Our models indicate that the most likely explanation for the outflow
clumps is that they are pre-existing, meaning $only$ that their
density structure is, at least partly, formed prior of the advent of
the outflow. This does $not$ exclude the general explanation that the
outflow clumps are mainly made of swept-up ambient gas and that
therefore the clumps are an indication of the episodic nature of the
outflows.

\item[4.]
It is probable that, with the advent of the outflow, not only the
temperature of the clumps increases and reaches the one observed, but
the clumps also undergo a period of non-dissociative shock (and
therefore high temperatures, $\sim$ 1000 K).

\item[5.]The rich chemistry of the clumps observed along CB3, and L1157, seems to be a
consequence of a pre-existing density enhancement (either uniform or
already in clumps) and of its interaction with the outflow. The
latter, most likely, shocks and accelerate the gas, and possibly, if
episodic, induces its clumpiness. This is indicated by the high
abundance of methanol and some of the sulphur-bearing species. In fact
these molecules are formed by a combination of freeze out and surface
reactions, and shocked chemistry: both most efficient when the outflow
compresses already dense material at its passage.
\item[6.] 
We find that it is not possible for the outflow clumps to have a
uniform high density - a density gradient is needed in order to
account for the observed emission of most species.
\item[7.] 
CO, CH$_3$OH, CS and SO$_2$ are most likely emitted from the lower
density components, while SO and OCS come from an intermediate density
component. This chemical stratification supports the findings of
Bachiller et al. (2001).
\item[8.] 
At late times ($t > 10,000$ yr) H$_2$CO is $always$ over-abundant by
at least a couple of orders of magnitudes.  This discrepancy is
similar to that found for the clumps ahead of Herbig-Haro objects
(Viti et al. 2003): these objects however do not share a common
chemistry and we find no chemical reason why the abundance of H$_2$CO
predicted by our models should be much larger than apparently
observed. A possible explanation may be that the observed clumps are
smaller than the size implied by the observations. In fact, when
computing the theoretical column densities for smaller sizes (see
Section 3.3.4), the match between theory and observations improves.

\end{enumerate}

In conclusion, we suggest that interferometric observations of outflow
clumps closer to us than CB3 are performed. This may reveal the real
structure of these clumps and help constrain the models.

\section*{Acknowledgements}
The authors are indebted to Prof D. A. Williams for useful
conversations about clumpiness and for a critical reading of this
revised version.  SV and MB thank the Italian Space Agency for
financial support. SV acknowledges individual financial support from a
PPARC Advanced Fellowship. The authors are very grateful to the
referee for comments and suggestions that substantially improved the
manuscript.

\label{lastpage}

\begin{thebibliography}{}
\bibitem[\protect\citeauthoryear{}]{}Arce H. G., \& Goodman A. A., 2001, ApJ 554, 132 
\bibitem[\protect\citeauthoryear{}]{}Arce  H. G., \& Goodman A. A., 2002, ApJ 575, 928  
\bibitem[\protect\citeauthoryear{Bachiller}{1996}]{}Bachiller R., 1996, ARA\&A 34, 111 
\bibitem[\protect\citeauthoryear{}]{}Bachiller R., Mart\'{\i}n-Pintado J., Fuente A., 1991, A\&A 243, L21 
\bibitem[\protect\citeauthoryear{}]{}Bachiller R., P{\' e}rez Guti{\' e}rrez M., Kumar M.~S.~N., Tafalla M., 2001, A\&A 372, 899 
\bibitem[\protect\citeauthoryear{}]{}Bergin E. A., Neufeld D. A., Melnick G. J., 1999, ApJ 499, 777   
\bibitem[\protect\citeauthoryear{}]{}Bergin E.~A., Alves J., Huard T., Lada C.~J., 2002, ApJ 570, L101  
\bibitem[\protect\citeauthoryear{}]{}Bourke T.L., Garay G., Lehtinen K.K., et al., 1997, ApJ 476, 781
\bibitem[\protect\citeauthoryear{}]{}Charnley S.~B., 1997, ApJ 481, 396 
\bibitem[\protect\citeauthoryear{}]{}Codella C,, \& Bachiller R., 1999, A\&A 350, 659 
\bibitem[\protect\citeauthoryear{}]{}Dutrey A., Guilloteau S., Bachiller R., 1997, A\&A 317, L55
\bibitem[\protect\citeauthoryear{}]{}Falle S. A. E. G., \& Hartquist T. W., 2002, MNRAS 329, 195
\bibitem[\protect\citeauthoryear{}]{}Garrod R. T., Williams D. A., Hartquist T. W., Rawlings J. M. C., Viti S., 2003, A\&A, submitted 
\bibitem[\protect\citeauthoryear{}]{}Hatchell J., Thompson M.~A., Millar T.~J., MacDonald G.~H., 1998, A\&A 338, 713 
\bibitem[\protect\citeauthoryear{}]{}Hatchell J., Viti S., 2002, A\&A 381, L33  
\bibitem[\protect\citeauthoryear{}]{}Hester J. J., Stapelfeldt K. R., Scowen P. A., 1998, AJ 116, 372 
\bibitem[\protect\citeauthoryear{}]{}Le Teuff Y. H., Millar T. J., Markwick A. J., 2000, A\&AS 146, 157 
\bibitem[\protect\citeauthoryear{}]{}Millar T. J., Farquhar P. R. A., Willacy K., 1997, A\&AS 121, 139 
\bibitem[\protect\citeauthoryear{}]{}Millar T. J., \& Hatchell J., 1998, Faraday Discussions n. 109,``Chemistry and Physics of Molecules and Grains in Space'', p. 15 
\bibitem[\protect\citeauthoryear{}]{}Morata O., Girart J. M., Estalella R., 2003, A\&A 397, 181 
\bibitem[\protect\citeauthoryear{}]{}Oppenheimer M., \& Dalgarno A., 1974, ApJ 187, 231 
\bibitem[\protect\citeauthoryear{}]{}Rawlings J. M. C., Hartquist T. W., Menten K. M., Williams D. A., 1992, MNRAS 255, 471 
\bibitem[\protect\citeauthoryear{}]{}Richer J.S., Shepherd D.S., Cabrit S., Bachiller R., Churchwell E., 2000, in Protostars and Planets IV, ed. Mannings V., Boss A.P., Russell S.S., University of Arizona Press,p. 867 
\bibitem[\protect\citeauthoryear{}]{}Rowan-Robinson M., 1980, ApJS 44, 403 
\bibitem[\protect\citeauthoryear{}]{}Ruffle D. P., Hartquist T. W., Caselli P., Williams D. A., 1999, MNRAS 306, 691 
\bibitem[\protect\citeauthoryear{}]{}Tafalla M., Myers P. C., Caselli P., Walmsley C. M., Comito C., 2002, ApJ 569, 815 
\bibitem[\protect\citeauthoryear{}]{}van Dishoeck E. W.,1998, in The Molecular Astrophysics of Stars and Galaxies, ed. T. W. Hartquist \& D. A. Williams (Clarendon Press, Oxford), 53 
\bibitem[\protect\citeauthoryear{}]{}Viti S., \& Williams D. A., 1999, MNRAS 310, 517 
\bibitem[\protect\citeauthoryear{}]{}Viti S., Girart J. M., Garrod R., Williams D. A., Estalella R., 2003, A\&A 399, 187 
\end{thebibliography}
\end{document}